# Two-Dimensional Transition Metal Dichalcogenides with a Hexagonal Lattice: Room Temperature Quantum Spin Hall Insulators

**(Submitted to PRB on 14 Oct. 2015)**

Yandong Ma[†,*], Liangzhi Kou[‡], Xiao Li[∥], Ying Dai[§], and Thomas Heine[†,⊥,*]

[†] Department of Physics and Earth Science, Jacobs University Bremen, Campus Ring 1, 28759 Bremen, Germany

[‡] School of Chemistry, Physics and Mechanical Engineering Faculty, Queensland University of Technology, Garden Point Campus, QLD 4001, Brisbane, Australia

[∥] Department of Physics, University of Texas at Austin, Austin, TX 78712, USA

[§] School of Physics, Shandong University, Shandanan Str. 27, 250100 Jinan, People's Republic of China

[⊥] Wilhelm-Ostwald-Institut für Physikalische und Theoretische Chemie, Universität Leipzig, Linnéstr. 2, 04103 Leipzig, Germany.

*Corresponding author: myd1987@gmail.com (Y.M.); thomas.heine@uni-leipzig.de (T.H.)

So far, several transition metal dichalcogenides (TMDCs) based two-dimensional (2D) topological insulators (TIs) have been discovered, all of them based on a tetragonal lattice. However, in 2D crystals, the hexagonal rather than the tetragonal symmetry is the most common motif. Here, based on first-principles calculations, we propose a new class of stable 2D TMDCs of composition $MX_2$ (M=Mo, W, X=S, Se, Te) with a hexagonal lattice. They are all in the same stability range as other 2D TMDC allotropes that have been demonstrated experimentally, and they are identified to be practical 2D TIs with large band gaps ranging from 41 to 198 meV, making them suitable for applications at room-temperature. Besides, in contrast to tetragonal 2D TMDs, their hexagonal lattice will greatly facilitate the integration of theses novel TI states van-der-Waals crystals with other hexagonal or honeycomb materials, and thus provide a route for 2D-material-based devices for wider nanoelectronic and spintronic applications. The nontrivial band gaps of both $WSe_2$ and $WTe_2$ 2D crystals are 198 meV, which are larger than that in any previously reported TMDC-based TIs. These large band gaps entirely stem from the strong spin-orbit coupling strength within the *d* orbitals of Mo/W atoms near the Fermi level. Our findings will significantly broaden the scientific and technological impact of both 2D TIs and TMDCs.

**I. Introduction**

The recent discovery of topological insulators (TIs), materials that act as insulators in their bulks yet support quantized gapless surface or edge states, has unleashed tremendous interest in condensed matter physics and material science owning to their rich physics and potential applications [1-3]. Charge carriers in such surface or edge states are helical Dirac fermions, which, different from conventional Dirac fermions, behave as massless relativistic particles with an intrinsic spin locked to its translational momentum because of time-reversal symmetry [4]. This allows the unique possibility of realizing coherent spin transport without heat dissipation [5,6]. Particularly in two-dimensional (2D) TIs, also called quantum spin Hall (QSH) insulators, the electrons at the edge can only move along two directions with opposite spins and thus insensitive to nonmagnetic chemical and structural edge modification, making 2D TIs better suited for potential dissipationless device applications [7]. Unfortunately, the realization of 2D TIs in experiment is challenging. Up to now, despite plenty of three-dimensional (3D) TIs being identified experimentally [8,9], the only experimental confirmation of 2D TIs was reported in HgTe/CdTe and InAs/GaSb/AlSb quantum well systems in an extreme experimental condition below 10 K [10-12], due to the tiny nontrivial bulk band gap and other complex factors in dealing with edge states. Therefore, searching for more feasible room temperature 2D TIs especially in commonly used materials is of great importance for their practical utilization. To this end, various strategies to produce 2D TI with large band gap have been followed [13-17].

Among all the predicted and synthesized 2D crystals, transition metal dichalcogenides (TMDCs), such as $MX_2$ with M=(Mo, W) and X=(S, Se, Te), are of particular appealing because of their unique material properties and potential applications in various electronic and optical devices [18-23]. Possessing strong intrinsic spin orbit coupling (SOC) [18-21], $MX_2$ 2D crystals have the potential to present QSH effect at room temperature, as well to be desirable to combine topological insulator with nanoelectronic devices. Indeed, several exploratory works have already been carried out in this realm. For instance, $MX_2$ 2D crystals with a distorted octahedral structure (termed as T′-$MX_2$ – as we will discuss exclusively 2D crystals with a single layer we will omit the stacking index in all crystal phases) were recently predicted to be new 2D TIs and topological field-effect transistors based on van der Waals heterostructures of T′-$MX_2$ and 2D dielectric layers were proposed [24]. In addition to the T′ phase, which possesses a rectangular lattice, 2D $MX_2$ crystals with a square lattice (termed as S-$MX_2$ or also $MX_2$ Haeckelite) were predicted to be QSH insulators [25-29]. Nevertheless, among all the previously reported 2D TIs as well as 2D trivial insulators, hexagonal, rather than the tetragonal, lattices are most commonly found in nature. This gives rise to the fundamental question: Is it possible to realize the QSH effect in hexagonal 2D $MX_2$ crystals, and if so, will it be observable at room temperature? Establishing these expectations successfully would not only enrich material science but also may lead to an untold number of applications.

In the present study, using *ab initio* simulations, we will show that the QSH effect can indeed exist in hexagonal 2D $MX_2$ crystals (H′-$MX_2$). Inspired by recent experimental and theoretical works on the stable grain boundary structures of TMDC 2D crystals [30-32], we predict a new family of H′-$MX_2$ (M=Mo, W, X=S, $Se_2$, Te). By combing phonon dispersion calculations with finite temperature Born-Oppenheimer molecular dynamics (BOMD) simulations, we demonstrate the dynamical and thermal stability of these H′-TMDC structures except for unstable H′-$MoTe_2$. We will further reveal all these systems are 2D TIs with a sizable bulk band gap, ranging from 41meV to 198meV, thus satisfying room-temperature spintronic applications. In particular, the nontrivial bulk band gap of 198 meV in H′-$WSe_2$ and $WTe_2$ is the largest value among all the reported TMDC-based 2D TIs.

**II. Computational Details**

Density functional theory (DFT) based first-principles calculations are performed using the projector augmented wave (PAW) method [33] as implemented in the VASP code [34,35]. The exchange-correlation energy is treated using the generalized gradient approximation (GGA) in the form of Perdew-Burke-Ernzerhof (PBE) functional [36]. The cutoff energy is 500 eV. Sufficient vacuum of more than 18 Å is used along the z direction, i.e., perpendicular to the 2D sheet, to avoid spurious interaction among the periodic images. The Brillouin zone (BZ) is sampled for integrations according to the Monkhorst−Pack scheme [37], with grids of 9×9×1 and 11×11×1 $k$-points, respectively, for geometric optimization and self-consistent calculations. All the structures are fully relaxed until the residual forces on each atom are smaller than 0.01 eV/Å. SOC is taken into account in terms of the second-variational procedure with the scalar relativistic eigenfunctions as basis set [38]. For the *ab initio* BOMD simulations, a canonical ensemble is simulated using the algorithm of Nosé. The cutoff energy is 400 eV. We adopt a relatively large supercell of 2×2 unit cells, with lattice parameter larger than 18 Å×18 Å. The phonon frequencies are calculated by using DFPT method [39] as implemented in the PHONOPY code [40].

**III. Results and Discussion**

In the commonly studied phase (namely the H phase) of $MX_2$ 2D crystals, the lattice is built entirely by the six-membered rings consisting of hetero-element M-X bonds. In addition to the regular six-membered rings, other types of rings have also been observed at the grain boundaries in H-$MX_2$ 2D crystals [30-32]. Inspired by these stable grain-boundary structures, we build a new phase of $MX_2$ 2D crystals; that is, H′-$MX_2$. Top and side views of the optimized crystal structure of the H′-$MX_2$ 2D crystals are shown in **Figure 1**. The structure crystallizes in the space group $D_{6h}$ and each unit cell contains six transition metal (M) and twelve chalcogen (X) atoms. From the side view, as shown in **Figure 1(b)**, a sandwich configuration is observed, where the M layer is inserted between two layers of X atoms via strong ionic-covalent bonding. While the top view shows that, in addition to the six-membered rings, the new phase is also composed of four- and twelve-membered rings, forming a 2D hexagonal lattice. Interestingly, we note that this new structure resembles the structure of the recent experimentally identified monolayer C2N-h2D [41]. By replacing the $C_4N_2$ rings with the four-membered M-X rings, the structure of H′-MX$_2$ can be obtained. The optimized in-plane lattice constant is found to be 8.815, 9.147, 9.712, 8.803, 9.134, and 9.696 Å, respectively, for WS$_2$, WSe$_2$, WTe$_2$, MoS$_2$, MoSe$_2$, and MoTe$_2$ 2D crystals. The relative energies of the five phases of MX$_2$ 2D crystals are listed in **Table 1**, and demonstrate that the thermodynamic stability of the H′ phase is comparable to those of the corresponding T phase. From **Table 1**, we can see that the total energy difference between H′ phase and H phase is a little large. However, it should be noted that T phase is also energetically less favored, even so, T phase 2D crystals have been obtained in many experiments [42-44]. Furthermore, recently, T phase 2D crystals have already been used as a cocatalyst in a silicon-based photoelectrochemical cell for hydrogen evolution [44] and as platforms for selective gas sensing [45]. Considering the fact that the total energies of H′ phase are comparable to that of T phase (the energy difference between H′ phase and T phase is only 23, -22, -43, 2, 3 and -1 meV, respectively, for WS$_2$, WSe$_2$, WTe$_2$, MoS$_2$, MoSe$_2$ and MoTe$_2$ 2D crystals), it is reasonable to expect that our proposed H′ structure could be found at least with the same probability as T configuration. Interestingly, T' phase is more stable than T phase, which is in good agreement with the previous work [24].

**Table 1**. Relative total energies (meV/atom) of different phases of MX$_2$ 2D crystals. The corresponding detailed atomic configurations are shown in **Figure S1**.

|    | WS$_2$ | WSe$_2$ | WTe$_2$ | MoS$_2$ | MoSe$_2$ | MoTe$_2$ |
|----|--------|---------|---------|---------|----------|----------|
| H  | 0      | 0       | 0       | 0       | 0        | 0        |
| T  | 296    | 258     | 189     | 280     | 235      | 172      |
| T′ | 180    | 90      | -29     | 184     | 110      | 14       |
| S  | 326    | 268     | 181     | 286     | 243      | 183      |
| H′ | 319    | 236     | 146     | 300     | 238      | 171      |

Phonon dispersions of H′-MX$_2$ 2D crystals are presented in **Figure 1(d)** and **Figure S2**. For WS$_2$, WSe$_2$, WTe$_2$, MoS$_2$ and MoSe$_2$, all branches over the entire Brillouin zone have positive frequencies, thus indicating that these structures are dynamically stable and that their stability does not depend on the substrate. In contrast to the typical linear dispersion of the acoustic modes near the Γ point, the lowest acoustic band (the out-of-plane acoustic mode) shows a q$^2$ dispersion, which is a characteristic feature of the phonon dispersion in layered materials, it was, for example, observed experimentally in graphite, boron nitride, and gallium sulfide [46]. In contrast, H′-MoTe$_2$ 2D crystal is instable, a result of an imaginary frequency around the M point (see **Figure S2**), resulting predominantly from the out-of-plane vibrations. According to **Figure 1(d)** and **Figure S2**, it is intriguing to note that, when moving from X=S to Te, the phonon branches in H′-WX$_2$ 2D crystal are shifted to lower frequencies. This finding mainly stems from that the mass of the X atoms increases with the sequence of S< Se< Te [47]. Similar behavior is also observed in H′-MoX$_2$ 2D crystals.

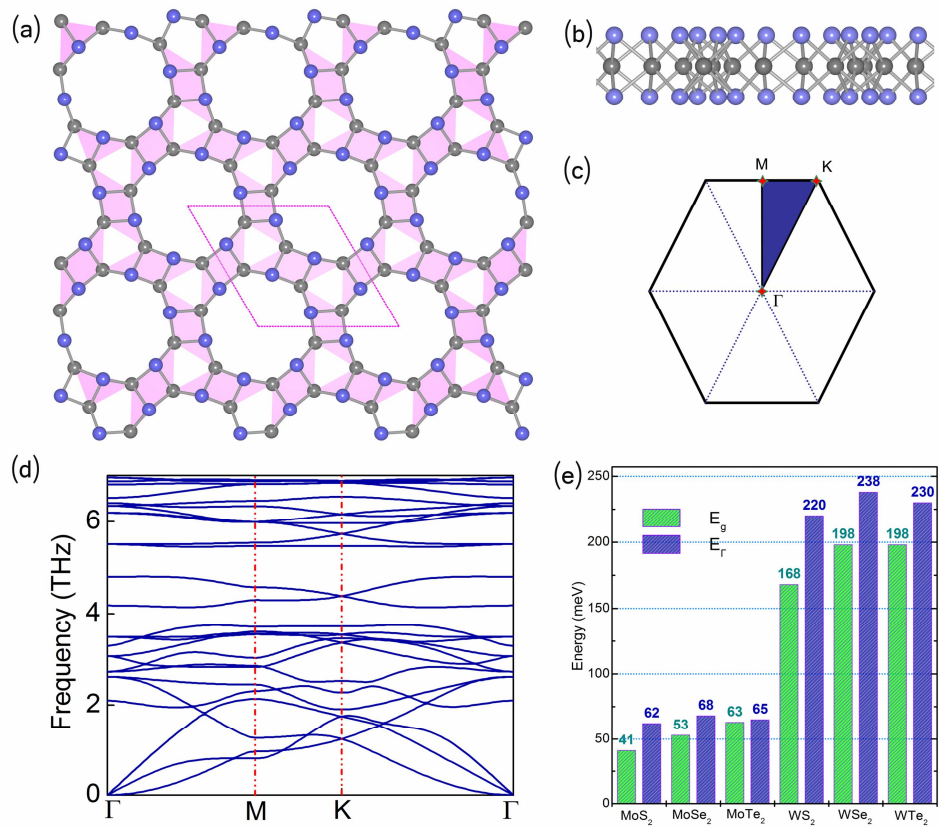

**Figure 1**. Atomic configuration of H′-MX$_2$ 2D crystals from the (a) top view and (b) side view. Gray and blue balls denote M and X atoms, respectively. The rectangle marked by pink dashed lines

denotes a unit cell. (c) Brillouin zone of H′-MX$_2$ 2D crystals with high-symmetry points. (d) Phonon spectra of H′-WS$_2$ 2D crystals. Phonon dispersions of the other H′-MX$_2$ 2D crystals are found in **Figure S2**. (e) The fundamental band gap ($E_g$) and the SOC-induced band gap ($E_\Gamma$) at the Γ point of H′-MX$_2$ 2D crystals.

A more rigorous test to probe the stability of WS$_2$, WSe$_2$, WTe$_2$, MoS$_2$ and MoSe$_2$ 2D crystals are *ab initio* BOMD simulations. Snapshots of atomic configurations of these 2D crystals at the end of the BOMD simulation are shown in **Figure 2**, **Figure S3** and **Figure S4**. After annealing at 300K and 500 K for 2.5 ps with a time step of 1 fs, neither structure disruption nor structure reconstruction was found to occur in all these systems. Moreover, we find that geometry optimizations of several molecular dynamics snapshots for these 2D crystals result in their corresponding original structures. To further confirm the thermal stability of these systems, we extend the annealing time of BOMD simulations from 2.5 ps to 10 ps. The corresponding results are presented in **Figure S5** and **S6**. We can see that, even after annealing at 300K and 500K for 10 ps, there is still neither structure disruption nor structure reconstruction in all these systems. These results support that the free-standing 2D crystals of H′-MX$_2$ (with the exception of WTe$_2$) are thermally stable and this new phase is separated by high-energy barriers from other local minima on the potential energy surface. Based on the above discussion, and taking into account that various allotropes of TMDC 2D crystals of similar energy have been reported from experiment, we thus can conclude that, although the most stable phase of WS$_2$, WSe$_2$, WTe$_2$, MoS$_2$ and MoSe$_2$ 2D crystals is H, the H′ phase could also be stable at or even above the room temperature. In the following, even though H′-MoTe$_2$ 2D crystal is unstable, its electronic properties are still investigated for comparison.

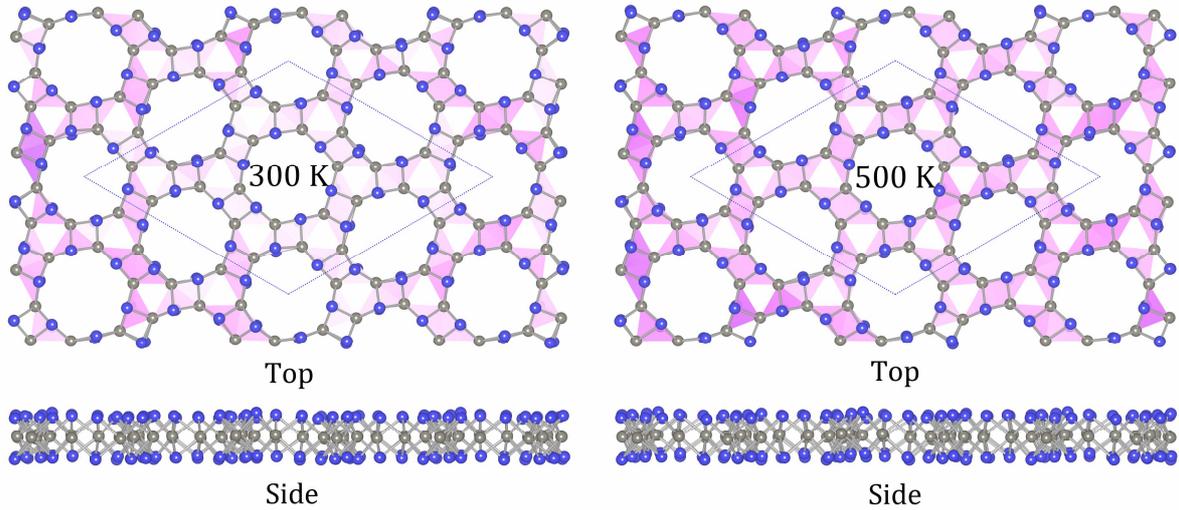

**Figure 2**. Snapshots of atomic configurations of H′-WS$_2$ at the end of the molecules dynamic simulation (with annealing time of 2.5 ps). The simulated supercells are marked by blue dashed lines, and their corresponding temperature is also denoted. Gray and blue balls denote W and S atoms, respectively.

Since M atoms possess strong intrinsic SOC [18,19], a fundamental precondition to induce nontrivial topology after ensuring its robust structural stability, it is interesting and important to reveal whether the H′-MX$_2$ 2D crystals exhibit TI features. To this end, it is instructive to first examine their electronic properties without the inclusion of SOC. The typical electronic band structures of H′-WS$_2$ and H′-WSe$_2$ 2D crystals without SOC are shown in **Figure 3(a)** and **(b)**, respectively, while the corresponding band structures of the other systems are plotted in **Figure S7**. Interestingly, the H′-MX$_2$ 2D crystals are gapless semiconductors, or alternatively semimetals, with the highest valence band (VB) and lowest conduction band (CB) degenerate at the Γ point, in sharp contrast to the insulating H-MX$_2$ 2D crystal and metallic T-MX$_2$ 2D crystal. Away from the Γ point, the CB and VB are well separated from each other. Such particular scenario for the band structure without SOC is a strong indication of the existence of a topological phase. In H′-MX$_2$ 2D crystals, the near-gap bands are mainly contributed by the *d* orbitals of the M atoms, with feeble contribution from the *p* orbital of the X atoms. Under the crystal field of H′-MX$_2$ 2D crystals, the *d* orbital of the M atom splits into three groups: $d_{z^2}$, $d_{xz/yz}$, and $d_{xy/x^2-y^2}$. Electronic orbital analysis manifests that the CB near the Γ point mainly originates from the M-$d_{z^2}$ orbitals, while the VB near the Γ point mainly originates from the hybridized M-$d_{xy/x^2-y^2}$ orbitals, as illustrated by the orbit-resolved band structures in **Figure 3(c)**. What is special here is that the state situating at the Fermi level is dominated by the mixed orbitals of M-$d_{z^2}$ and M-$d_{xy/x^2-y^2}$ due to the degeneracy of the VB and CB.

To properly address of the electronic band structure of H′-MX$_2$ 2D crystals, one needs to go beyond calculations in scalar relativistic approximation and include SOC. Electronic band structures of H′-MX$_2$ 2D crystals with the involvement of SOC are plotted in **Figure 3** and **Figure S7**. Clearly, the SOC interaction lifts the degeneracy of the CB and VB at the Γ point and shifts the CB (VB) upwards (downwards), resulting in an insulating phase with a large band gap, see **Figure 3** and **Figure S7**. Here we label the SOC-induced band gap at the Γ point as $E_\Gamma$ to be distinguished from the global band gap $E_g$. Since any gap opening at the touching points must originate from SOC, $E_\Gamma$ provides an ideal measurement for the SOC strength. As shown in **Figure 1(e)**, the global band gaps of WS$_2$, WSe$_2$, WTe$_2$, MoS$_2$, MoSe$_2$, and MoTe$_2$ 2D crystals are 168, 198, 198, 41, 53, and 63 meV, respectively. Their large bulk band gaps can effectively protect the edge current against the interference of thermally activated carriers in the bulk. Such large band gaps mainly stem from the strong SOC within the *d* orbitals of the M atoms. In previous work [48,49], in order to obtain TIs with visible SOC-induced nontrivial gap, metal atoms are deposited on the surface to hybridize the *p* orbitals around the Fermi level with the *d* orbitals of metal atoms. Notably, unlike these works, it appears to be the better alternative to design materials where the states around the Fermi level are governed by *d* orbitals.

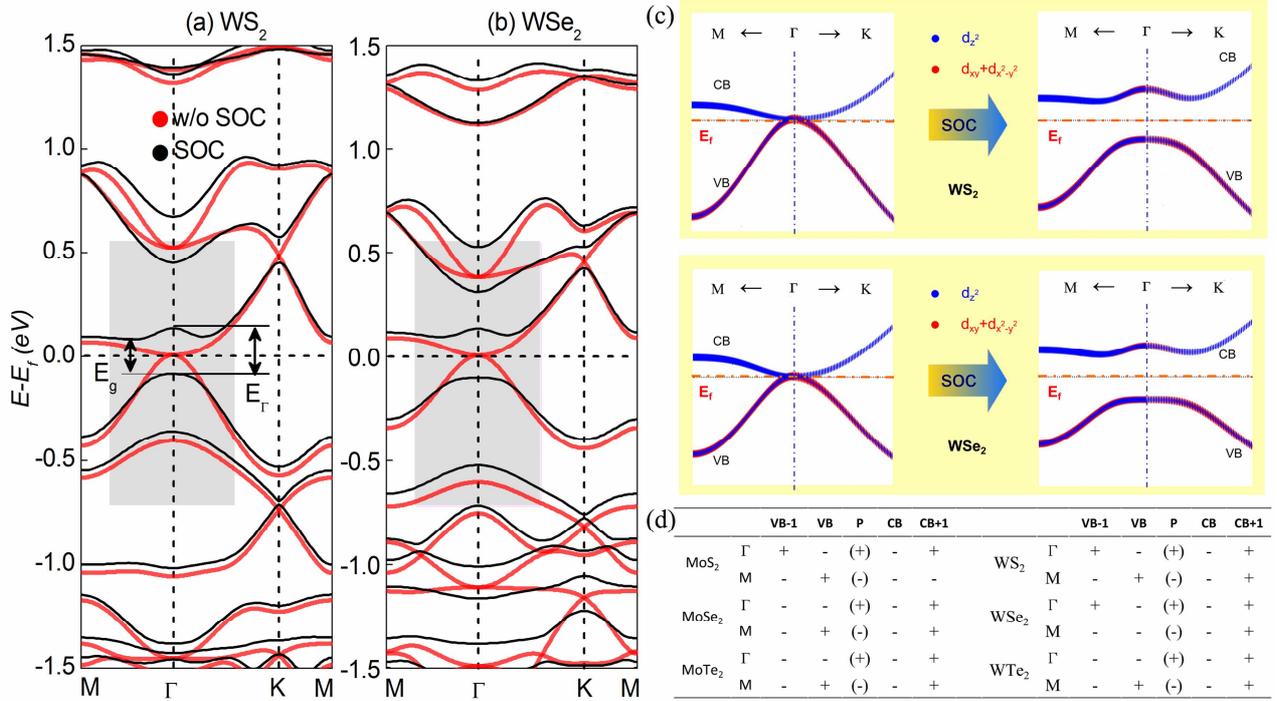

**Figure 3**. Electronic band structures of (a) H′-WS$_2$ and (b) H′-WSe$_2$ 2D crystals without and with SOC. Fermi levels are set to zero. (c) The W-$d_{z2}$, W-$d_{xy/x2-y2}$ orbital projected character of CB and VB around the Γ point for H′-WS$_2$ and H′-WSe$_2$ 2D crystals without and with SOC. (d) The parities of VB-1, VB, CB and CB+1 at the Γ point for MX$_2$; the products of the occupied bands at the Γ and M points are also listed in the brackets.

To understand the SOC-induced band splittings in H′-MX$_2$ 2D crystals, we still take H′-WS$_2$ and H′-WSe$_2$ 2D crystals as examples. We investigated the evolution of the orbital-resolved CB and VB near the Γ point with SOC. The corresponding band structures are illustrated in **Figure 3(c)**. Recalling that, when excluding SOC, CB and VB around the Γ point are mainly constituted by M-$d_{z2}$ and M-$d_{xy/x2-y2}$ orbitals, respectively, and they are well separated from each other except for the touching point at Γ, it is therefore supposed that the inclusion of SOC would just simply remove the degenerate point and divide M-$d_{z2}$ orbitals from M-$d_{xy/x2-y2}$ orbitals. However, this is not the case. According to **Figure 3(c)**, around the Γ point, the inclusion of SOC will first mix the M-$d_{z2}$ and M-$d_{xy/x2-y2}$ orbitals and then split them. Consequently, different from the case without SOC, both the CB and VB near the Γ point are mainly composed of not only M-$d_{z2}$ but also M-$d_{xy/x2-y2}$ orbitals, whereas for CB and VB away from the Γ point, they are still dominated by M-$d_{z2}$ and M-$d_{xy/x2-y2}$ orbitals, respectively. Such band feature suggests a nontrivial topological phase in H′-MX$_2$ 2D crystals. Moreover, the SOC-induced *W*-shape CB and roughly *M*-shape VB provide another sign of a nontrivial phase [50,51].

To firmly identify the nontrivial topological phase in H′-MX$_2$ 2D crystals, we have implemented the direct calculation of Z$_2$ topological invariants (ν) by evaluating the parity eigenvalues of the occupied states at all time-reversal invariant momentum (TRIM) points of Brillouin zone [52]. The first Brillouin zone of H′-MX$_2$ 2D crystal is plotted in **Figure 1(c)**. There are four TRIM points: Γ and three M points. The products of the parities of the Bloch wave functions for the occupied bands at the Γ and M points are shown in **Figure 3(d)**. For all the H′-MX$_2$ 2D crystals, the product of parity eigenvalues at Γ point is +1, whereas it is -1 at M point, thus yielding a nontrivial topological invariant ν=1. Accordingly, all these hexagonal H′-MX$_2$ 2D crystals are indeed 2D TIs. More complete information regarding the parity eigenvalues of the second-highest valence band (VB-1), VB, CB and second-lowest conduction band (CB+1) at Γ and M points are displayed in **Figure 3(d)**. It is important to notice that both VB and CB at the Γ point comprising M-$d_{z2}$ and M-$d_{xy/x2-y2}$ orbitals show odd parity. Therefore the underlying mechanism is that, as SOC is taken into account, the Fermi level separates one odd parity state from another odd parity state and a nontrivial topological nature is then established. During this process, no parity exchange is induced.

Given these results, we believe that the QSH effect can be observed at room temperature in H′-MX$_2$ 2D crystals, in particular for H′-WSe$_2$ and H′-WTe$_2$ with their nontrivial bulk band gap of 198 meV, which are so far the largest band gaps of all the reported TMDC-based 2D TIs [24-27], and which correspond to roughly eight times the thermal energy of room temperature. Aside from the large band gap, another virtue of H′-MX$_2$ 2D crystals as a practical QSH system is their hexagonal lattice. Note that the previously reported TMDC-based 2D TIs all display tetragonal lattices, while hexagonal symmetry is the most common motif in the field of 2D materials. For facilitating the integration of theses new TI states in 2D-material-based devices, such for topological field-effect transistors made from a combination of alternating layers of 2D materials, hexagonal lattice is a natural advantage as it provides a better registry. These large bulk band gaps, along with the hexagonal lattice, provide strong motivation for realizing QSH effect in H′-MX$_2$ 2D crystals.

It is interesting to compare the values of the SOC strength E$_\Gamma$ with those of the global band gaps E$_g$ of these H′-MX$_2$ 2D crystals. From **Figure 1(e)**, we can see that, although the global band gaps of the H′-MoX$_2$ (MoS$_2$, MoSe$_2$, MoTe$_2$) 2D crystals are significantly different from each other (41 meV, 53 meV, 63 meV), the values of SOC strength of them are roughly comparable (68 meV, 62 meV, 65 meV). A similar trend is observed in the H′-WX$_2$ 2D crystals. On the other hand, since the SOC strength increases with the sequence of S < Se < Te, the values of the SOC strength E$_\Gamma$ of H′-MX$_2$ 2D crystals are supposed to increase with the sequence of MS$_2$ < MSe$_2$ < MTe$_2$. To explore the origin of this apparent inconsistency, we need to focus on states around the Fermi level at the Γ point in view of the fact that its splitting caused by SOC directly determines the E$_\Gamma$. As we demonstrated above, the states around the Γ point are mainly constituted by M-$d_{z2}$ and M-$d_{xy/x2-y2}$ orbitals, whereas the contribution from the X atoms can almost be ignored. This means that the E$_\Gamma$ of H′-MX$_2$ 2D crystals only depends on the SOC strength of M-$d_{z2}$ and M-$d_{xy/x2-y2}$ orbitals and the SOC strength of X atoms is not contributing. Thus, it is understandable that the values for E$_\Gamma$ of MoX$_2$ H′-MoX$_2$ (WX$_2$) 2D crystals are roughly comparable with each other. This conclusion is further corroborated by the fact that the values of the SOC strength of all H′-MoX$_2$ 2D crystals are much smaller than those of all H′-WX$_2$ 2D crystals, in accordance with the fact that the SOC strength of Mo atom is much smaller than that of the W atom.

At last, we comment on the possible routes for synthesizing H′-MX$_2$ 2D crystals. Since the total energies of H′ phases are comparable to that of T phases, probably H′ phase can be realized by co-opting the similar ways used to prepare T phase [42-44]. One possible approach to realize the H′-MX$_2$ 2D crystals is to grow the H′-MX$_2$ 2D crystals on a substrate which interacts weakly with the films. The substrate growth method is well developed now. An alternative way is the colloidal chemical synthesis strategy, which can also be employed to prepare H′-MX$_2$ 2D crystals. Such two well-developed methods, which are effective in preparing the energetically comparable T-MX$_2$ 2D crystals, both could in principle realize H′-MX$_2$ monolayers with high crystallinity. Besides, the H' phase may coexist with the robust H phase within the same sample. Under such condition, the nontrivial topological performance of the H' phase may be affected. Therefore, a further study concerning this should be performed in further works.

## IV. Conclusion

In conclusion, we report the discovery of a novel stable phase of MX$_2$ (MoS$_2$, MoSe$_2$, WS$_2$, WSe$_2$) 2D crystals with hexagonal symmetry. Their stability is firmly confirmed by combing phonon dispersion calculations with finite temperature molecular dynamic simulations. Importantly, we find that all these systems are promising candidates for 2D TIs with a sizable bulk band gap ranging from 41 meV to 198 meV, thus promising for room-temperature applications. Specially, the nontrivial bulk band gaps of H′-WSe$_2$ and H′-WTe$_2$ 2D crystals even reach up to 198 meV. These results represent a significant advance in developing 2D TIs based on TMDCs.

## Supporting Information

The atomic configurations of different phases of MX$_2$ 2D crystals; phonon spectra of H′-WSe$_2$, H′-WTe$_2$, H′-MoS$_2$, H′-MoSe$_2$, and H′-MoTe$_2$ 2D crystals; snapshots of atomic configurations of H′-WSe$_2$, H′-WTe$_2$ H′-MoS$_2$ and H′-MoSe$_2$ 2D crystals at the end of the molecules dynamic simulation; electronic band structures of H′-WTe$_2$, H′-MoS$_2$, H′-MoSe$_2$, and H′-MoTe$_2$ 2D crystals without and with SOC. This material is available free of charge via the Internet at.

## Acknowledgement

Financial support by the European Research Council (ERC, StG 256962) and the Taishan Scholar Program of Shandong are gratefully acknowledged.

## Note

The authors declare no competing financial interest.